\documentclass[aps,prd,twocolumn,groupedaddress,nofootinbib]{revtex4-1}
\usepackage{graphicx}

\begin{document}


\title{Eternal Inflation\\with Arrival Terminals }

\author{Henry Stoltenberg and Andreas Albrecht}
\affiliation{University of California at Davis;
Department of Physics\\
One Shields Avenue;
Davis, CA 95616\\
}



\begin{abstract}
We analyze the cosmological role of terminal vacua in the string
theory landscape, and point out that existing work on this topic makes
very strong assumptions about the properties of the terminal vacua. We
explore the implications of relaxing these assumptions (by including
``arrival'' as well as ``departure'' terminals) and demonstrate that
the results in earlier work are highly sensitive to their assumption
of no arrival terminals.  We use our discussion
to make some general points about tuning and initial
conditions in cosmology. 
\end{abstract}


\maketitle

\section{Introduction}

Cosmology seeks to provide a history of our universe,
describing the dynamical evolution from its initial 
state to the world we observe today.  The question of fine tuning of
initial conditions in cosmology was one of the original motivations
for cosmic inflation theory~\cite{Guth:1980zm}, and it remains a
difficult problem~\cite{Albrecht:2014eaa}.  Thanks to our everyday
experience with the 2nd law of thermodynamics, it seems very natural
to accept initial conditions that have low entropy.  However, low
entropy initial conditions are necessarily fine tuned.  In this paper
we make some specific points about how tuning assumptions can slip in
essentially unnoticed into cosmological discussions. 

One challenging related issue is the
possibility that our observations do not result from a natural
cosmological evolution but instead are the result of a thermal
(or other) fluctuation. The observations of this fluctuation-based
observer (or ``Boltzmann Brain''~\cite{Albrecht:2004ke}) need not
reflect the universe or its physical laws. Cosmological theories which
predict we are more likely to be these Boltzmann Brains
instead of ``normal'' observers are unsatisfactory and such a prediction is usually
considered grounds for rejection of the theory in question. 

In an eternally inflating multiverse in a landscape there are many
different types of vacua with different physics. Throughout the history
of this ``multiverse'' transitions among these different vacua will
take place, typically via bubble transitions which form ``pocket
universes'' in the new vacuum. According
to standard treatments\footnote{See~\cite{Albrecht:2014eaa} for an alternative}, a landscape with
only vacua with positive cosmological constants (de Sitter-like vacua)
would reach an equilibrium fixed point regardless of initial 
conditions.  This equilibrium would be dominated by Boltzmann Brain
observers.   

Page~\cite{Page:2006dt} showed that if our vacuum decays at a fast
enough rate then Boltzmann Brain type observers would be in the
minority. Bousso and Freivogel ~\cite{Bousso:2006xc} as well as
Linde~\cite{Linde:2006nw} applied this concept to a local viewpoint
allowing de Sitter-like vacua to decay to ``terminal'' vacua (Anti-de
Sitter-like vacua which collapse to a singularity). 
Bousso et
al.~\cite{Bousso:2008hz,Bousso:2011aa} and Harlow et
al.~\cite{Harlow:2011az} 
have taken this further, showing that introducing ``terminal'' vacua can remove the
problematic equilibrium fixed point.
A terminal vacuum will crunch to a singularity in a finite time and thus no longer
participate in any further transitions. If the rate of tunneling out
of a vacuum that supports normal observers to terminals is significantly greater
than the rate for producing Boltzmann Brains, the Boltzmann Brain
problem would be resolved.  Furthermore, the state of the universe
will approach a special steady state that is determined by the values
of the various tunneling rates. 

Many assumptions about the nature of terminals and the initial state
of the universe have gone into this
description. These assumptions can be explored by taking a closer look
at how the terminals are described: A bubble with negative cosmological constant will crunch
at a rate faster than an unlikely up-tunneling which is why
transitions out of terminals are typically not
considered.\footnote{Transitions from AdS vacua to dS has been
  considered by Garriga and Vilenkin ~\cite{Garriga:2012bc} by
  removing the singularity from AdS space allowing for a bounce during
  the crunch with an arrow of time arising from special initial
  conditions. We do not use this particular mechanism for transitions
  in this paper.} In addition 
the physics
of the Anti-de Sitter (AdS) pocket universe as it nears a singularity is not well
understood, so one could not even apply something like Coleman-de Luccia
tunneling rates as one approaches the singularity.  

But if we believe that physics is fundamentally
unitary one expects that the crunching spacetime continues to evolve
after the crunch.  There is no clear theory of what this evolution
would look like, and no strong reason to expect it would be described
in terms of the local field theoretic ideas on which the rest of the
picture is based. Nonetheless, if one believes the theory is unitary,
the evolution must continue.  For this paper we simply consider a ``completion'' of the
theory where we say there is some hidden part of the Hilbert
space in which the unitary evolution continues on the other side of
the singularity. Since we cannot access it directly and have no clear
picture what that evolution is like, we just call it the ``hidden
Hilbert space'' (HHS).  Again evoking unitarity, one needs to consider the time reverse of the AdS
crunch. Such evolution would describe an allowed process in the
theory where a subsystem would start in the HHS and pop out through a
time reversed singularity into the time reverse of an AdS
crunch. This
time reversed picture could be extended further to
describe time reversed tunneling from AdS out into another vacuum. In
this paper we refer to the usual terminals considered
in~\cite{Linde:2006nw,Bousso:2008hz,Bousso:2011aa,Harlow:2011az} as ``departure terminals'', and the time
reverse of these (which can describe transitions from AdS to dS vacua) as
``arrival terminals''. Note that since departure terminals can occur
at any time in the cosmological evolution, unitarity would
allow arrival terminals to appear at any time as well. 

\begin{figure}[b]
\includegraphics[width = 3.4in]{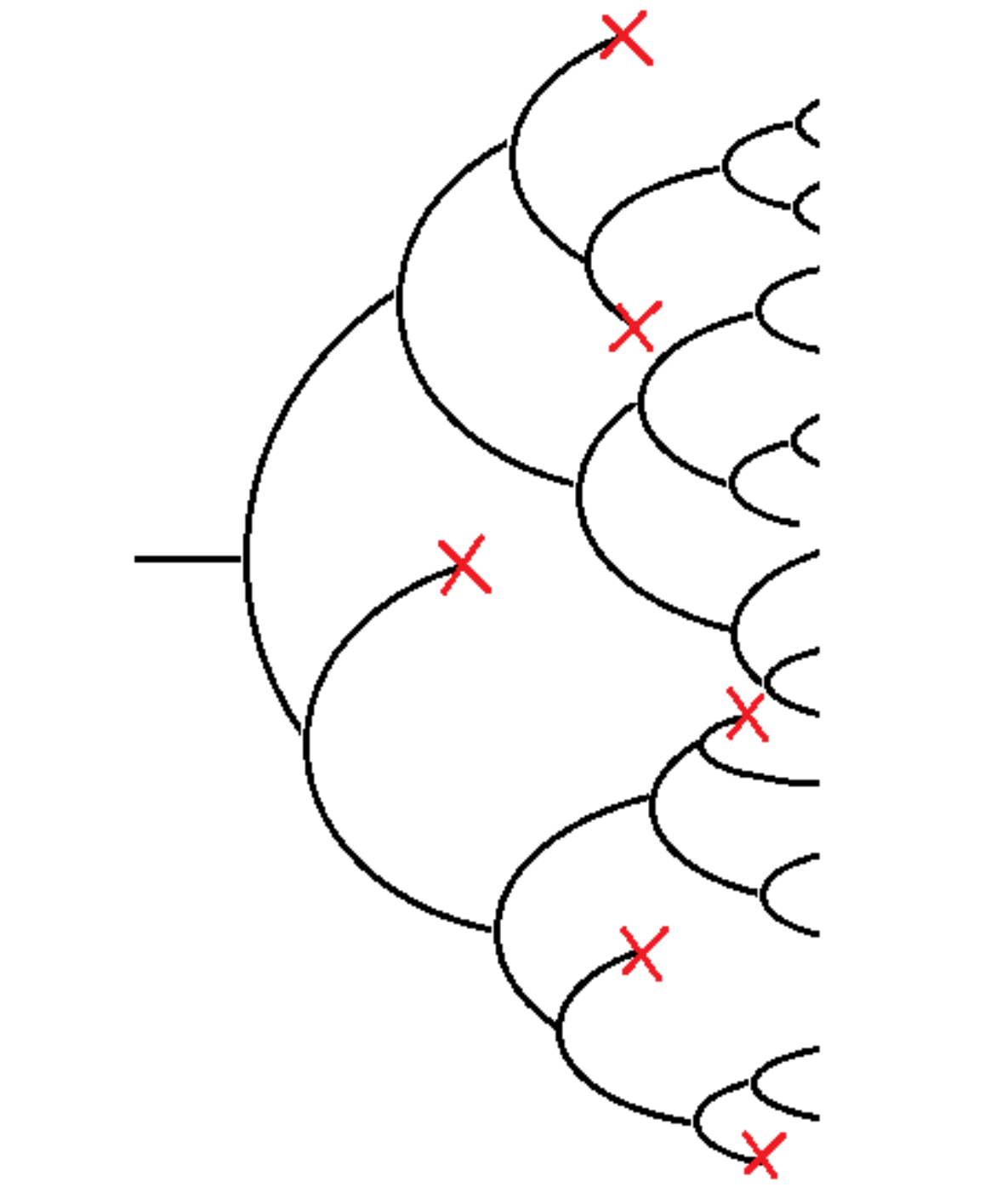}
\caption{\label{fig:treedepart} The tree like structure represents a
discrete model of eternal inflation like the one presented in
\cite{Harlow:2011az}. The tree follows the causal future of an inflating bubble
with time flowing from left to right. Branching of the tree represents
the exponential expansion of de Sitter space as well as bubble
nucleation, with separated branches falling out of causal
contact. Terminal vacua are marked with an $X$, where we no longer
follow their evolution once they have crunched. } 
\end{figure}

\begin{figure}[b]
\includegraphics[width = 3.4 in]{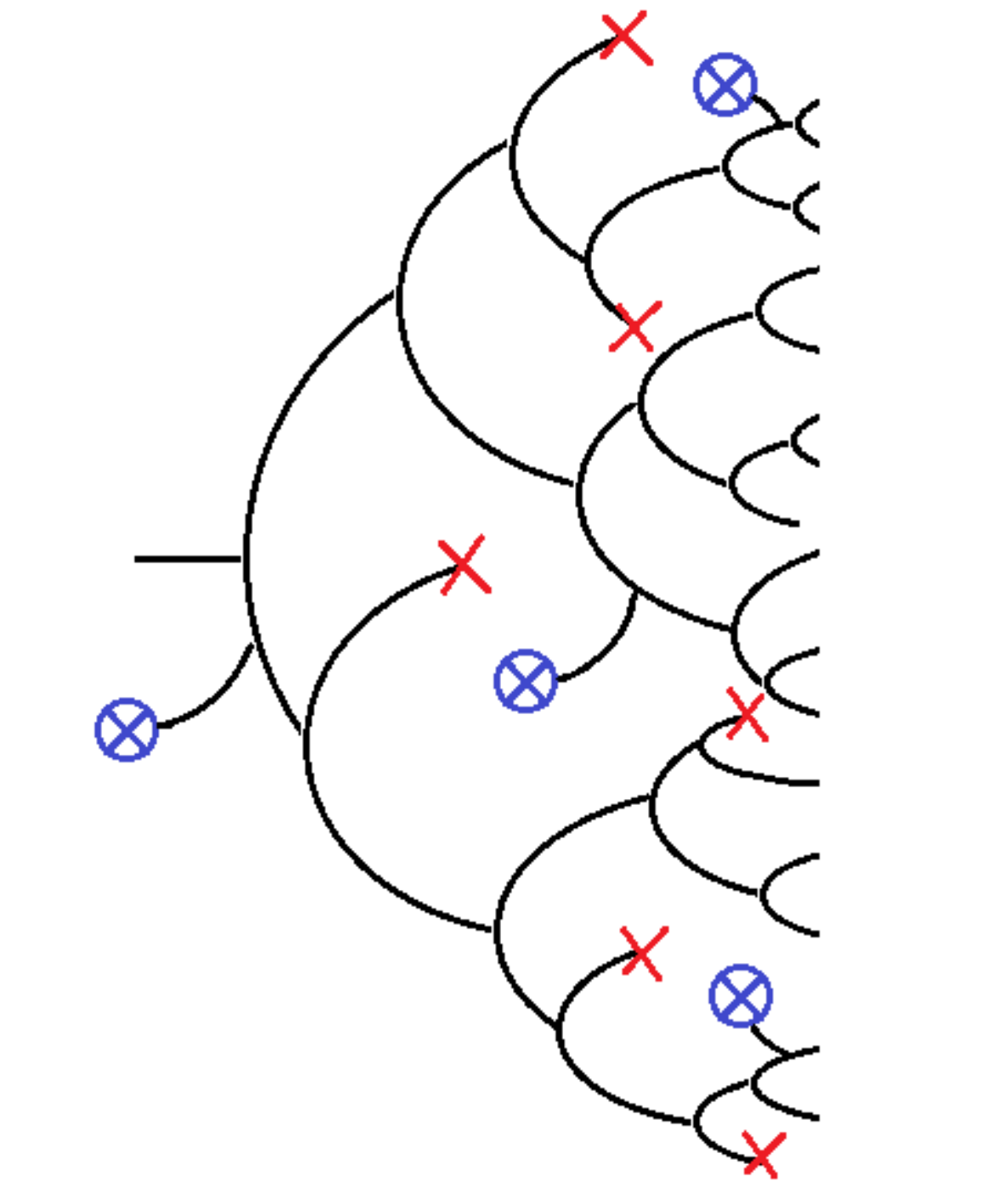}
\caption{\label{fig:treearrive} We modify the tree model shown in
  Fig.~\ref{fig:treedepart} by allowing
arrival terminals represented by $\bigotimes$'s. For unitary theories
such arrival terminals represent allowed processes, and the choice to
exclude them is effectively a choice about initial conditions.}
\end{figure}

This work is different from past treatments in that the rate of arrival terminals
popping in is not described by known physics like the Coleman-de
Luccia tunneling.  Instead, it is related to unknown physics in the
HHS which feeds into the time reverse of the tunneling process.  Thus
we can no longer assume the rate of transitions from
AdS to dS to be negligible. Comoving volume and probability
that leaks into departure terminals in principle can return from
arrival terminals. In this paper, we ignore AdS bubbles which arrive
then crunch without tunneling to a dS vacuum and use an effective rate
of transitioning from arrival terminal to dS type vacuum in
Eqn~\ref{dfiA}. 
Figure~\ref{fig:treedepart} shows a standard picture of
landscape evolution with only departure terminals present (meant to
correspond to Fig.~4 from~\cite{Harlow:2011az}),  
while our Fig.~\ref{fig:treearrive} adds in arrival terminals as well.

Terminals with zero vacuum energy which evolve asymptotically 
to Minkowski space at late times were
also considered in~\cite{Harlow:2011az}
(following~\cite{Susskind:2007pv} these are called  ``hats'').  Our
arguments about allowing time reversed 
behavior also apply to hat-type terminals (in this case the
arrival terminals would correspond to special states coming in from
infinitely early times and ultimately tunneling to a different vacuum
state). For the purposes of our discussion we lump these hats (and
their time reverse) together with the AdS terminals. While the special
assumptions and properties of the AdS crunch discussed here do look
very different from the asymptotic Minkowski behaviors in the zero
vacuum energy, they can still be lumped together to make our main points.

There could exist some special subset of initial
quantum states of the multiverse which for the entirety of it's evolution will completely avoid sectors
of the HHS that create arrival terminals. We frame past treatments such as the analysis in~\cite{Harlow:2011az} with only
departure type terminals as specially chosen ``initial conditions'' in our
picture.  We put ``initial conditions'' in quotes
to acknowledge that the complicated dynamics we are contemplating for
``beyond the AdS crunch'' may well not lend themselves to a simple
initial time slice picture for the entire system.  But still, the
presence or absence of various arrival terminals certainly counts as
the technical equivalent to assumptions about initial conditions. The term initial may be misleading however since
it does not only describe the initial volume of AdS type vacua but also states in the HHS which may lead to future AdS arrivals not on the initial time slice.

In this paper we show that the special steady state attractor found
in~\cite{Harlow:2011az} depends heavily on their choice of an
arrival-free state.  The steady state found
in~\cite{Harlow:2011az} is a feature put in ``by hand''
via the choice of initial state, rather than a feature determined
intrinsically by the tunneling rates alone, as it might first appear. We
demonstrate this by making different assumptions about arrival
terminals in a simple toy model and show that these lead to different
steady state attractors.
Thus we disagree with the claim in~\cite{Harlow:2011az} that their attractor ``depends on the
existence of an initial condition but not on its
details''~\footnote{Technically this disagreement depends on whether
  their assumption of no arrivals should be attributed to initial
  conditions or something else (a purely semantic question).  Here we emphasize the non-semantic
result that the attractor depends sensitively on assumptions about the state which
are put in by hand.}.

The topics of equilibrium and the arrow of time are closely tied to
our discussion. Without {\em any} terminals, the system is understood to reach
equilibrium (maximum entropy) which means that the arrow of time only
appears as a transient, before the equilibrium sets in.  It is the
equilibrium behavior that leads to the Boltzmann Brain problem.  In
the picture with only departure 
terminals proposed in~\cite{Harlow:2011az} the system does exhibit an arrow of time, and would never
reach equilibrium.  The reason for this behavior is the specially tuned choice
of initial state (specifically, the assumption that arrival terminals
are absent for all eternity). In this respect the arrow of time
identified in~\cite{Harlow:2011az} originates in the usual way, as the
result of a special choice of low entropy initial conditions
(see~\cite{Penrose:1980ge,Albrecht:2002uz,Albrecht:2014eaa} for discussions of this
point in the context of cosmology). 

This paper is organized as follows: Section~\ref{formalism} reviews
the standard formalism used in earlier work to analyze the no-arrivals
case, and then extends that formalism to include arrival
terminals. Section~\ref{toymodel} applies the formalism to a toy
model, and demonstrates the dependence of the steady state attractor
on the assumptions about arrival terminals. Section~\ref{summary}
develops our discussion of the toy model further, presenting specific
solutions that illustrate our main point.  We also summarize our main
points and conclusions in Sect.~\ref{summary}.

\section{Transition Rates and Steady States}
\label{formalism}
\subsection{No arrival terminals}
This section reviews the formalism developed in~\cite{Garriga:2005av}
(and utilized in~\cite{Harlow:2011az,Bousso:2008hz,Bousso:2011aa})
as applied only to departure terminals.  One can define
a dimensionless transition rate from vacuum $j$ to vacuum $i$,  as 
\begin{equation}
\kappa_{ij}=\frac{4\pi\Gamma_{ij}}{3H_j^4}
\end{equation}
Where $\Gamma_{ij}$ is the bubble nucleation rate per physical volume 
for bubbles of type $i$ in bubbles of type $j$. Here $H_j$ is the expansion
rate of bubble $j$. The $\Gamma_{ij}'$'s could be given by the
Coleman-de Luccia tunneling rates in the thin wall approximation but
their exact form isn't important for this discussion.  

In the absence of arrival terminals, in a
particular measure we can define probabilities from the fraction of
comoving volume occupied by vacuum $i$, $f_i$. Since we are interested
in observers that form in a space with a positive cosmological
constant, we will only keep track of volume fractions that correspond
to de Sitter-like vacua (which are non-terminal). The equation that
describes how these volume fractions evolve with time is given by: 
\begin{eqnarray}
\frac{df_i}{dt}=\sum_{j}\kappa_{ij}f_j-\kappa_if_i
\label{dfi}\\
\kappa_i\equiv\ \kappa_i^D + \sum_{j}\kappa_{ji}.
\end{eqnarray}
The quantity $\kappa_i$ is defined to be the total transition rate out of vacuum $i$
which includes rates to transition into other de Sitter type vacua,
$\kappa_{ji}$ and rates to transition to departure terminals,
$\kappa_i^D$. Since these equations do not keep track of the volume
fraction that is lost to departure terminals, $f$ would have to be
renormalized to define a probability. 

This equation can be rewritten using a matrix, $M$:
\begin{eqnarray}
\frac{df_i}{dt}=\sum_{j}M_{ij}f_j 
\\
\label{eq:M}
M_{ij}\equiv\kappa_{ij}-\kappa_i\delta_{ij}
\end{eqnarray}
Solutions to this are of the form:
\begin{equation}
f_i=\sum_{l}s_i^le^{-q^lt}
\end{equation}
where $-q^l$ are the eigenvalues of $M$ and $s_i^l$ are the
corresponding eigenvectors. Since
the $f_i$'s as we have defined them do not include terminal vacua
there is an effective leak of volume fraction. This property and the
assumption that each de Sitter vacuum is accessible from each other in
one or more transitions result in all eigenvalues being negative and
non-zero. The least negative eigenvalue is non-degenerate(shown in the
appendix of \cite{Garriga:2005av}), which is called the dominant
eigenvalue. The dominant eigenvalue (along with the corresponding
eigenvector) will dominate late time behavior. We denote
the dominant eigenvector/eigenvalue without the $l$ superscript and write our
late time solution as: 
\begin{equation}
\label{eq:fi}
f_i=s_ie^{-qt}
\end{equation}
This attractor solution is seen in~\cite{Harlow:2011az} as giving rise to an emergent arrow of
time, as probability flows trough the dominant steady state given by
$q$ and out into the departure terminals. 

\subsection{Adding arrival terminals}
We now extend the above formalism to include arrival terminals.  The
timescale for tunneling out of an AdS terminal vacuum before it has 
crunched is much longer than the timescale for nearing the singularity
and approaching (poorly understood) Planck scale physics. Transitions of the sort which
could be described with the typical Coleman de-Luccia rates would be
highly sub-dominant. Here we focus on allowing transitions that are
simply the time reverse of departure terminal crunches (in the AdS
case, or alternatively the time reverse of ``hats'' in the zero vacuum
energy case).  We do not presume to understand
these processes at a technical level (since we don't have a
detailed theory of the crunch) but we expect such processes to be
allowed if the full theory is unitary. 

To allow transitions from ``arrival terminals'' we 
introduce an additional term in Eqn.~\ref{dfi} to get:
\begin{equation}
\frac{df_i}{dt}=\sum_{j}\kappa_{ij}f_j-\kappa_if_i+\kappa_i^Af^A(t)
\label{dfiA}
\end{equation}
This equation is constructed such that it mirrors the form of
Eqn.~\ref{dfi} which has a clear space-time description. Here, the
quantity $f^A(t)$ looks like 
an effective comoving volume fraction for the arrival terminals in
which case $\kappa_i^A$ would be like a rate to transition from
arrival terminal to vacuum $i$. However we want to emphasize that the
for HHS states we do not necessarily expect to have a space-time interpretation. We  
leave  $\kappa_i^Af^A(t)$ as an effective rate that could vary with
time.  

The function $f^A(t)$ could in principle be anything due to our lack of knowledge
of the physics of AdS vacua once they have crunched, and similar open
questions about ``incoming hats''. We just consider
a simple case where in some sense the arrival terminal vacua are being
depleted as transitions occur. We will characterize this depletion
with a simple decaying exponential with some time constant, R: 
\begin{equation}
f^A(t)=e^{-Rt}
\end{equation}
We emphasize that this approach does not consider a fully generic
initial state, but just one that is a bit more generic than one gets by assuming
no arrivals at all.   This choice is meant to easily fit methods of previous work while
demonstrating our main point (the sensitivity of the steady state to
choices about initial conditions). In the general case one would not expect to
even have an arrow of time or a steady state, but our modifications
are not sufficiently general to demonstrate that aspect.  In that
sense our modification keeps the ``temporal provincialism''~\cite{Dyson:2002pf} exhibited
in~\cite{Harlow:2011az}. 

\section{Toy Model}
\label{toymodel}
To make our main points we now apply our formalism to a simple toy
model. 
\begin{figure}[b]
\includegraphics[width = 3.4 in, trim=0in 2.5in 0in 2.5in]{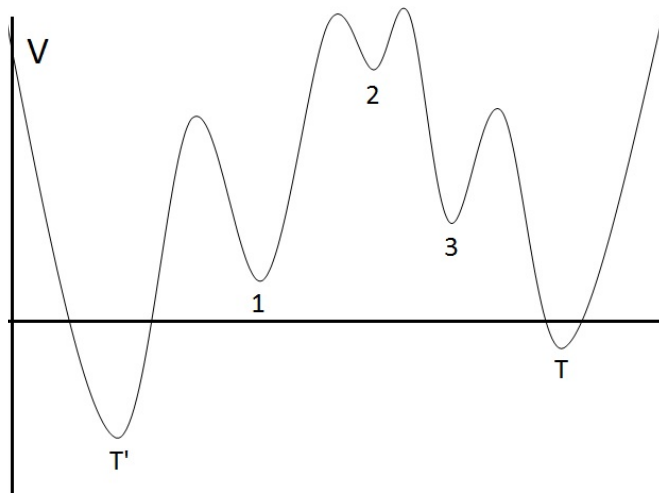}
\caption{\label{fig:landscape} A sketch of our toy model landscape. Vacua $2$, $3$ and $1$ (in
  order of decreasing $\Lambda$) are the only vacua with positive
  cosmological constant. Vacua $T$ and $T'$ are terminals and was
  assume are only accessible from vacua $1$ and $3$ respectively. }
\end{figure}
Our toy model landscape has three
de-Sitter vacua and two anti de-Sitter vacua very similar to the case considered
in~\cite{Bousso:2011aa}. The potential for this landscape is shown in
Fig.~\ref{fig:landscape}. The original purpose of this toy landscape
was to show how terminal vacua could resolve Boltzmann Brain problems in certain
cases by producing an attractor solution with the right properties. We
will show that allowing arrival terminals can change the properties of
the attractor solution, and can thus reintroduce the
Boltzmann Brain problem depending on the specific choices made about arrivals. 

Starting with the case of departure terminals but no arrival terminals, Eq.~\ref{eq:M} is:
\begin{eqnarray}
M_{ij}=
\left(
\begin{array}{ccc}
-\kappa_1 & \kappa_{12} & 0 \\
\kappa_{21} & -\kappa_2  & \kappa_{23} \\
0 & \kappa_{32} & -\kappa_3 
\end{array}
\right).
\\
\end{eqnarray}
Since in this model we have assumed vacuum $1$ has a much lower
$\Lambda$ than vacuum $2$, the probability of up tunneling from vacuum $1$
to $2$ is much lower than the total decay rate out of $1$. 
\begin{equation}
\epsilon \equiv \frac{\kappa_{21}}{\kappa_1} \ll 1
\end{equation}
To first order in $\epsilon$, the dominant eigenvalue ($Q$) and
eigenstate $\vec{s}$ are give by 
\begin{eqnarray}
q = \kappa_1 -
\frac{\epsilon(\kappa_3-\kappa_1)\kappa_{12}\kappa_1}{(\kappa_3-\kappa_1)(\kappa_2-\kappa_1)-\kappa_{32}\kappa_{23}}
\\ 
\vec{s}=
\left(
\begin{array}{c}
1 \\
\frac{\epsilon(\kappa_3-\kappa_1)\kappa_{1}\kappa_1}{(\kappa_3-\kappa_1)(\kappa_2-\kappa_1)-\kappa_{32}\kappa_{23}} \\
\frac{\epsilon\kappa_{32}\kappa_1}{(\kappa_3-\kappa_1)(\kappa_2-\kappa_1)-\kappa_{32}\kappa_{23}} .

\end{array}
\right)
\label{ToyqsNA}
\end{eqnarray}
For convenience, the dominant eigenvector, $\vec{s}$ is normalized
such that the entry for the longest lived vacuum is unity. 

Now let's modify this to include arrival terminals. For simplicity we
will only allow arrival terminals to tunnel into vacua $1$ and $3$. The
solution given by using Eqns.~\ref{ToyqsNA} in Eqn.~\ref{eq:fi} will
now be the homogeneous solution to Eqn.~\ref{dfi}.  Including the
contributions from arrival terminals will give
the general solution: 
\begin{equation}
f_i(t)=b(Cs_ie^{-qt}+\frac{\kappa_i^A}{\kappa_i-R}e^{-Rt})
\end{equation}
where $C$ is an arbitrary constant determined by initial conditions and $b$ is an overall normalization factor. 

In the case where $R << q$, the second term in the equation will
dominate at late times. The ratio of volume fractions are then
strongly governed by the arrival rates. 

\section{Discussion and Conclusions}
\label{summary}
In this paper we try to implement unitarity which we believe should be
fundamental in the full theory. That leads us to add a sector to the
Hilbert space (which we call the HHS) to account for evolution on the
other side of singularities. We do not assume the HHS has a space-time
description and suspect it may not. We have chosen to describe interactions
between the spacetime sector and the HHS through additional terms of
similar form to the volume fractions and transition rates used to
describe transitions among actual vacua. This choice is made for
convenience and does not describe the most general case. In our
formalism, from the point of view of the space-time sector the considerations
presented in this paper do not seem to alter certain known features
for the multiverse. In this view, the departure/arrival terminals may look
effectively like just another type of vacuum that dS vacua can transition to and
from. The result (discussed in~\cite{DeSimone:2008bq, Garriga:2005av, Harlow:2011az}) that the final attractor is independent of the
initial volume fractions and depends only on the transition rates is
unchanged. However, our point is that including information about all
the actual vacua in the landscape is not enough. One also has
to include terms to represent transitions from the HHS to the
space-time sector (the arrival terminals).  While in a certain formal
sense the earlier results carry over, our point is that the earlier
work has implicitly made assumptions about the HHS which have not been
justified and are not at all general.  In fact, one can imagine more
general scenarios which give arrival terminals that arrive more
randomly in the space-time sector (not
describable by the extended equations we have developed here) in which the HHS interactions with the
space-time sector prevent any attractor from appearing. Admittedly,
understanding which assumptions and scenarios are realistic depends on
achieving much greater clarity on the nature of the HHS than anyone
has at present. 

Even sticking to our extended equations (and despite formal similarities with earlier
work) we find that the inclusion of arrival terminals can dramatically change the 
evolution of the multiverse.  It can change the steady state
attractor that emerges and can significantly modify
the probability of forming Boltzmann Brains. Our toy model is adapted
from~\cite{Bousso:2011aa}, where it is
assumed that vacuum $3$ is the only vacuum capable of
supporting any type of observer. Comparing the probability of ordinary
observers to Boltzmann Brain observers is done using the dominant
eigenvector (or dominant history)  method, which evaluates the relative abundance of Boltzmann
Brains vs. legitimate observers in the steady state which emerges
after an initial transient.  
In the previous section we showed that with arrival terminals, the dominant
 eigenvector would be heavily dependent on the arrival rates. While we do
 not have a formalism like the Coleman-De Luccia tunneling rates to
 characterize the arrival rates, one still cannot make them disappear without
effectively making an assumption about either the initial quantum state or not allowing the theory to be unitary.

To give a concrete illustration on how the dominant eigenvector can be
affected, we numerically solved Eqn.~\ref{dfiA} for some arbitrarily chosen
transition rates that are characteristic of our toy model: 
 
 \begin{eqnarray}
M = 
\left(
\begin{array}{ccc}
 -0.01 & 0.03 & 0 \\
 0.003 & -0.055 & 0.002 \\
 0 & 0.025 & -0.02
\end{array}
\right)
\end{eqnarray}

\begin{figure*}[t]
\includegraphics [width = 5.5 in, trim=0in 2.5in 0in 2.5in] {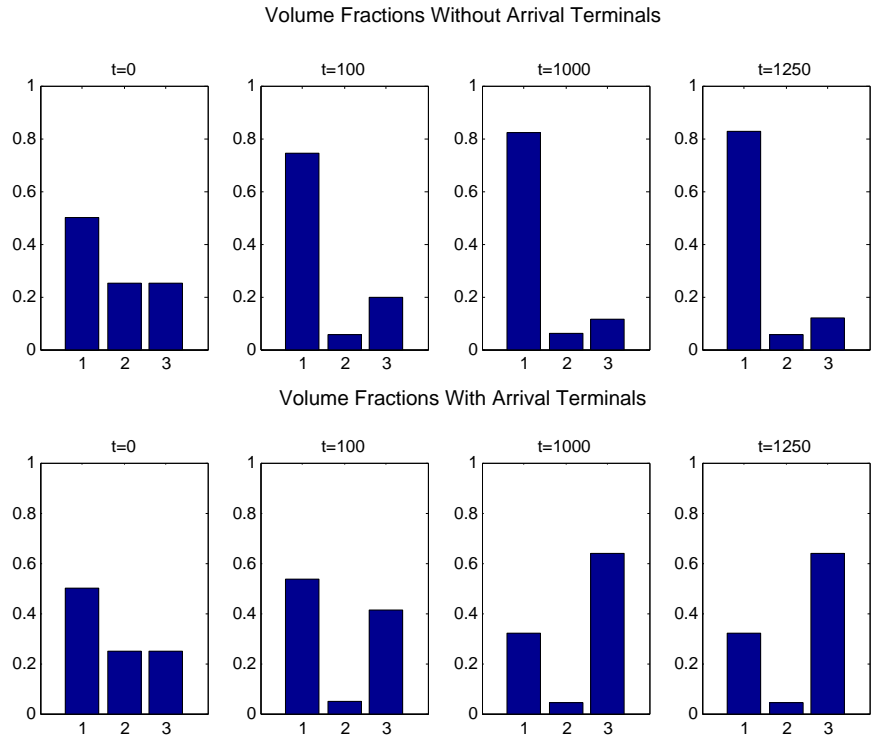}
\caption{\label{fig:volfrac} 
Renormalized volume fractions for de Sitter vacua $1$, $2$
and $3$ from our toy model at various times. The top row corresponds to
the scenario where the arrival rates are exactly zero and the bottom
row has arrival rates: $\kappa_A^1=0.0005$, $\kappa_A^2=0$,
$\kappa_A^3=0.004$ and time constant: $R=0.002$. Initially there is
transient behavior but by t=1000 both scenarios have reached the
dominant eigenvector and the renormalized volume fractions no longer
evolve with time. These results illustrate how the properties of the
dominant eigenvector depend on assumptions about arrival terminals. }
\end{figure*}

We considered the case where the arrival rates are exactly zero and
compared to the case where they are non-zero (specific values are
given in the caption to Fig.~\ref{fig:volfrac}). From the resulting volume
fractions shown in Fig.~\ref{fig:volfrac} we can see for 
initial conditions that differ only on the treatment of arrival
terminals, the dominant eigenvector achieved in the two 
scenarios are different. 
Without knowledge of what the arrival rates
should be, we are unable to make predictions on the dominant
eigenvector which prevents us from making predictions of the late-time
behavior of the theory. 

Our experience with a strongly expressed 2nd law (and the associated
low entropy past) in the world around
us can make it seem very natural to assume similar properties must be
true in the general cosmological case (a phenomenon aptly dubbed ``temporal provincialism''
in~\cite{Dyson:2002pf}). When exploring the fundamental origins of the
arrow of time in the universe one must take great care to avoid, or at
least clearly identify the extent to which we are allowing a temporally
provincial perspective.  We argue that the intuition that one is free
to exclude arrival terminals in landscape cosmologies, even while
allowing departure terminals, is an example of this temporally
provincial phenomenon.  This conclusion does not detract from
important points made in~\cite{Harlow:2011az,Bousso:2011aa}. Those
papers show the significant implications of imposing the (highly tuned) condition
that arrival terminals are absent, while departure terminals are
allowed. In particular, they show how adding only the departure terminals
can eliminate the Boltzmann Brain problem in models which would
otherwise have one. 

While we have focused our comments
on~\cite{Harlow:2011az,Bousso:2011aa}, our points are relevant more
generally, as reviewed in~\cite{Albrecht:2014eaa}.  For example
in~\cite{Carroll:2004pn}, the presence of an arrow of time is deeply
dependent on one's willingness to accept the assumption that the time
reverse of their tunneling process out of the background de Sitter
space is entirely absent. The de Sitter equilibrium
model~\cite{Albrecht:2014eaa} (which has many similarities with the
model presented in~\cite{Carroll:2004pn}) is very different in this
regard, since as a fundamentally equilibrium model the time reverse of
all processes are included, and their rates are given by the detailed
balance property of equilibrium systems. 

The origin of the thermodynamic arrow of time is a very important and
puzzling aspect of cosmology.  There
appear to be only two ways to incorporate it in our theories: One can put the arrow of time in by hand,
and accept the finely tuned  initial conditions this approach
requires, or consider an equilibrium model which requires exotic
dynamics to avoid the Boltzmann Brian
problem~\cite{Albrecht:2014eaa}. Our familiarity with a world that 
strongly exhibits an arrow of time can make it easy to overlook the
strong assumptions that go into the first approach. In this paper we
carefully analyze the tuning that appears
in~\cite{Harlow:2011az,Bousso:2011aa} via the the assumed absence of
arrival terminals. While the authors
of~\cite{Harlow:2011az,Bousso:2011aa} understand this tuning and
embrace it for the purposes of that work\footnote{R.~Bousso and
  L.~Susskind, private communication}, we still find it useful to call
out these aspects more explicitly than has been done in those papers.   We feel the sort 
of analysis presented here is helpful in clarifying our thinking about
the arrow of time in cosmology, and may lead us ultimately to a more
satisfactory understanding.

\acknowledgements
We thank R. Bousso and L. Susskind for 
helpful discussions.  This work was supported in part by DOE Grant
DE-FG02-91ER40674.

\bibliography{AA}

\end{document}